\documentstyle{mn}
\newif\ifAMStwofonts

\input{psfig.sty}

\ifoldfss
  \ifCUPmtlplainloaded \else
    \NewTextAlphabet{textbfit} {cmbxti10} {}
    \NewTextAlphabet{textbfss} {cmssbx10} {}
    \NewMathAlphabet{mathbfit} {cmbxti10} {} 
    \NewMathAlphabet{mathbfss} {cmssbx10} {} 
  \fi
  \ifAMStwofonts
    \ifCUPmtlplainloaded \else
      \NewSymbolFont{upmath} {eurm10}
      \NewSymbolFont{AMSa} {msam10}
      \NewMathSymbol{\upi}     {0}{upmath}{19}
      \NewMathSymbol{\umu}     {0}{upmath}{16}
      \NewMathSymbol{\upartial}{0}{upmath}{40}
      \NewMathSymbol{\leqslant}{3}{AMSa}{36}
      \NewMathSymbol{\geqslant}{3}{AMSa}{3E}

      \let\leq=\leqslant 
      \let\geq=\geqslant 
    \fi
  \fi
\fi 

\ifnfssone
  \newmathalphabet{\mathit}
  \addtoversion{normal}{\mathit}{cmr}{m}{it}
  \addtoversion{bold}{\mathit}{cmr}{bx}{it}
  \newmathalphabet{\mathbfit} 
  \addtoversion{normal}{\mathbfit}{cmr}{bx}{it}
  \addtoversion{bold}{\mathbfit}{cmr}{bx}{it}
  \newmathalphabet{\mathbfss} 
  \addtoversion{normal}{\mathbfss}{cmss}{bx}{n}
  \addtoversion{bold}{\mathbfss}{cmss}{bx}{n}
  \ifAMStwofonts
    \ifCUPmtlplainloaded \else
      \UseAMStwoboldmath
      \makeatletter
      \new@mathgroup\upmath@group
      \define@mathgroup\mv@normal\upmath@group{eur}{m}{n}
      \define@mathgroup\mv@bold\upmath@group{eur}{b}{n}
      \edef\UPM{\hexnumber\upmath@group}
      \new@mathgroup\amsa@group
      \define@mathgroup\mv@normal\amsa@group{msa}{m}{n}
      \define@mathgroup\mv@bold\amsa@group{msa}{m}{n}
      \edef\AMSa{\hexnumber\amsa@group}
      \makeatother
      \mathchardef\upi="0\UPM19
      \mathchardef\umu="0\UPM16
      \mathchardef\upartial="0\UPM40
      \mathchardef\leqslant="3\AMSa36
      \mathchardef\geqslant="3\AMSa3E

      \let\leq=\leqslant 
      \let\geq=\geqslant 
    \fi
  \fi
\fi 

\ifnfsstwo
  \DeclareMathAlphabet{\mathbfit}{OT1}{cmr}{bx}{it}
  \SetMathAlphabet\mathbfit{bold}{OT1}{cmr}{bx}{it}
  \DeclareMathAlphabet{\mathbfss}{OT1}{cmss}{bx}{n}
  \SetMathAlphabet\mathbfss{bold}{OT1}{cmss}{bx}{n}
  \ifAMStwofonts
    \ifCUPmtlplainloaded \else
      \DeclareSymbolFont{UPM}{U}{eur}{m}{n}
      \SetSymbolFont{UPM}{bold}{U}{eur}{b}{n}
      \DeclareSymbolFont{AMSa}{U}{msa}{m}{n}
      \DeclareMathSymbol{\upi}{0}{UPM}{"19}
      \DeclareMathSymbol{\umu}{0}{UPM}{"16}
      \DeclareMathSymbol{\upartial}{0}{UPM}{"40}
      \DeclareMathSymbol{\leqslant}{3}{AMSa}{"36}
      \DeclareMathSymbol{\geqslant}{3}{AMSa}{"3E}

      \let\leq=\leqslant 
      \let\geq=\geqslant 
    \fi
  \fi
\fi 

\ifCUPmtlplainloaded \else
  \ifAMStwofonts \else 
    \def\upi{\pi}
    \def\umu{\mu}
    \def\upartial{\partial}
  \fi
\fi

\begin{document}
\title[Morphological Classification of Spirals]{Morphological Classification of the OSU Bright Spiral Galaxy Survey}
\author[Whyte et al.]{
\parbox[t]{\textwidth}{L.F. Whyte$^1$, R.G. Abraham$^2$, M.R. Merrifield$^1$, 
   P.B. Eskridge$^3$, J.A. Frogel$^4$, and R.W. Pogge$^4$\\}
\vspace*{6pt}\\
$^1$School of Physics and Astronomy, University of Nottingham,
Nottingham, NG7 2RD, UK.\\ 
$^2$Department of Astronomy and Astrophysics, University of Toronto, 
60 St. George Street, Toronto, ON, M5S 3H8, Canada.\\
$^3$Department of Physics \& Astronomy, Minnesota State University, Mankato, MN 56003, USA \\
$^4$Department of Astronomy, The Ohio State University, 140 W. 18th Avenue, Columbus, OH 43210, USA. \\
} 
\date{Accepted 2002.
      Received 2002;
      in original form 2002 July 15}

\pagerange{\pageref{firstpage}--\pageref{lastpage}}
\pubyear{1994}

\maketitle

\label{firstpage}

\begin{abstract}

To quantify the distribution of bar shapes in spiral galaxies, we
have analysed 113 \textit{H}-band and 89 \textit{B}-band galaxy images from the Ohio
State University Bright Spiral Galaxy Survey.  Parameters measuring
bar shape and position along the Hubble sequence were obtained in
each waveband. Evidence was found for a bimodality in the
distribution of bar shape, implying that barred and unbarred
galaxies are not just the extrema of a single distribution, and that
any evolution between these two states must occur on a rapid
timescale.  Objective bar shapes measured in the \textit{H}-band were found
to be more closely related to visual classifications than \textit{B}-band bar
strengths, as the \textit{B}-band images are somewhat compromised by localised star
formation, especially in later-type systems.  Galaxies were found to
be more centrally concentrated in the infrared.  Later type galaxies
showed greater asymmetry in the optical than the infrared, presumably
again due to localised star formation, but on average the bar
strengths in the two bands were found to be the same.

\end{abstract}


\section{Introduction}

For over 75 years, bright galaxies have been classified using Hubble's
(1926) Tuning Fork.  Modifications and refinements have been made, but
the original scheme, defining ``early'' to ``late'' types by the size
of the bulge, and the smoothness and pitch angle of the spiral arms,
has proved to be a remarkably robust starting point.  Hubble's second
parameter, whether or not a galaxy contains a central bar, has also
proved a useful defining feature in describing galaxies' morphologies.
However, the existence of galaxies with differing degrees of
``barriness'' led de Vaucouleurs (1959) to introduce an intermediate
category, SAB, between strongly-barred SB galaxies and unbarred SA
galaxies.  This finer gradation raises the natural question of whether
galaxies really are naturally divided into barred and unbarred
systems, or whether some continuum exists.  This issue is clearly of
significance in trying to understand the still-disputed origins of the
bars found in a large fraction of galaxies [see, for example, Sellwood
(1999)].

In an attempt to address this question, Abraham \& Merrifield (2000)
made a quantitative assessment of the morphologies of galaxies in the
Frei et al.\ (1996) catalogue.  They discovered that if these galaxies
are plotted in a two parameter ``Hubble Space,'' with a measure of
position along the Hubble sequence on the $x$ axis and barriness on
the $y$ axis, then the galaxies naturally split into the classical
bifurcated tuning fork seen in Hubble's original work, with the barred
galaxies forming a remarkably tight sequence in this space.  The
bimodal nature of the barriness parameter then implies that barred and
unbarred galaxies do form distinct populations.  However, there were
two significant shortcomings in this analysis.  

First, the Frei et al.\ (1996) catalogue was not selected in a manner
well-suited to this type of analysis.  Indeed, since it was chosen to
contain a broad range of galaxy types, one might well imagine that the
most strongly barred galaxies were included to represent this type of
system, while galaxies with no central distortion in their isophotes
were selected to represent unbarred systems.  Thus, one could imagine
that the criteria upon which the catalogue was selected could introduce
exactly the type of bimodal distribution observed.

Second, the observations for the Frei et al.\ (1996) catalogue were all
at optical wavelengths.  Light in this part of the spectrum can be
dominated by relatively small amounts of recent star formation, so the
perceived morphology may not be representative of a galaxy's
underlying structure.  Hackwell \& Schweizer (1983), and more recently 
Block et al.\ (1994), have shown that a galaxy's
optical appearance can be totally different from its infrared
morphology, which is more representative of the bulk of the galaxy's
stellar distribution.  Thus, even such basic properties as whether or
not a galaxy contains a bar may be difficult to establish at optical
wavelengths.  Indeed, it is notable in Abraham \& Merrifield's (2000)
analysis that unbarred late-type galaxies, where contamination by
recent star formation is more of an issue, have tended to be
classified as intermediate ``SAB'' galaxies, perhaps reflecting the
difficulty in classifying the barriness of these systems. A further 
indication of the potentially misleading nature of optical
classifications of barriness comes from the simple statistic that at
visible wavelengths only $\sim 50-60\%$ of galaxies are classified as
barred (Sellwood \& Wilkinson 1993), whereas at infrared wavelengths
the fraction is $\sim 70\%$ (Mulchaey \& Regan 1997; Knapen et al. 2000; Eskridge et al.\ 2000).

In this paper, we address these issues by carrying out a morphological
analysis of data from the Ohio State University Bright Spiral Galaxy
Survey\footnotemark[1].
The galaxies in this sample have well defined selection criteria, and should prove representative of bright spiral galaxies in the local Universe, removing the first concern expressed above.
Further, galaxy images are available at both near infrared and optical
wavelengths, so the morphological properties can be determined using
the more robust infrared data, and tested against the morphologies
derived in the optical.

\footnotetext[1]{
The subset of the OSUBSGS used was the 
Early Data Release, available online at 
http://www.astronomy.ohio-state.edu/$\sim$survey/EDR/.}

The remainder of the paper is laid out as follows.  In
Section~\ref{sec:OSU}, the data set is briefly described.
Section~\ref{sec:class} describes the measures adopted to quantify
each galaxy's position in Hubble Space, while
Section~\ref{sec:results} presents the results of applying this
analysis.  Conclusions are discussed in Section~\ref{sec:disc}.

\section{The OSU Bright Spiral Galaxy Survey}\label{sec:OSU}

The sample of galaxies used in this investigation were taken from the
Ohio State University Bright Spiral Galaxy Survey (OSUBSGS).  205
galaxies were observed in B,V,R,J,H, and K bands at five different
observatories, using 1.2m - 2.4m telescopes.  A full description of
the survey can be found in Eskridge et al.\ (2002). The spiral
galaxies were selected for the survey from Third Reference Catalogue of
Bright Galaxies (RC3) (de Vaucouleurs et al. 1991) subject to the
constraints that they have a \textit{B}-band magnitude $B \leq 12$ and a
diameter $D \leq 6\rlap.'5$.

In this paper a total of 196 \textit{H}-band, and 166 \textit{B}-band galaxy images were
investigated.  However before the classification parameters could be
measured, the foreground stars in each image had to be removed.  Stars
were replaced with an average of the surrounding pixels.  All images
were closely examined for suitability and were rejected if they had
too many foreground stars; had bright stars near the centre; or had
large bright stars elsewhere causing saturation over a significant
portion of the galaxy image.

\begin{figure}
\begin{minipage}[b]{\linewidth}
\psfig{file=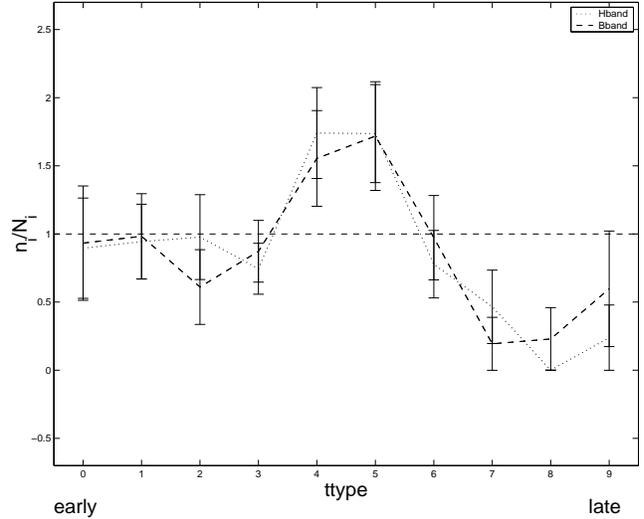,angle=270,width=\linewidth}
\end{minipage}
\caption{Distribution of galaxy Hubble types in the H and \textit{B}-band subsamples of OSUBSGS, in comparison with the RC3. $N_{i}$ is the normalised number of galaxies in the RC3, while $n_{i}$ is the normalised number of galaxies in the OSUBSGS subsamples, so for an unbiased sample $n_{i}/N_{i} = 1$.}
\label{rc3}
\end{figure}

As it is almost impossible to determined whether a galaxy that lies
close to edge-on is barred on the basis of photometry alone, we follow
Abraham et al. (1999) and only use galaxies whose axis ratios imply an
inclination $i < 60^{o}$.  In the \textit{B}-band images, galaxies with strong
dust or star formation features (e.g.\ NGC~2442), could not be
satisfactorily fitted by any kind of ellipse-fitting model, and were
therefore also rejected. In total 89 \textit{B}-band and 113 \textit{H}-band images were
found to be suitable, with a total of 72 galaxies having images in
both bands.  Fig.~\ref{rc3} shows that the sub-sample is slightly deficient of very late-type systems (due to the rejection of heavily dust-obscured galaxies), but is otherwise representative of the nearby bright spiral galaxy population found in the RC3.

\section{Classification Parameters}\label{sec:class}

In order to classify the galaxies, we adopt the framework described in
detail in Abraham \& Merrifield (2000). This methodology defines a
two-dimensional Hubble Space where the $x$-coordinate measures
position along the early-to-late sequence, while the $y$-coordinate
measures, in a quantitative way, the degree to which a galaxy is
barred.  This parameter space provides a quantitative framework for
investigating Hubble's Tuning Fork: if the original scheme were strictly
valid, we might expect to find galaxies separated into one-dimensional
barred and unbarred sequences, with weakly barred galaxies perhaps
lying at the lower edge of the barred distribution.

We assume that the distribution of light corresponds to a thin disk
that is intrinsically axisymmetric at large radii, and define the
centre of galaxy as the pixel with the maximum flux.  A quantitative
measure of asymmetry, $A$, was obtained by subtracting the image of
the galaxy from a version that had been rotated through 180 degrees
(Abraham et al. 1996).  We define the early-to-late ($x$) axis using
the central concentration parameter, $C$, defined in Abraham et al.\
(1994) and closely related to the parameter defined by Doi, Fukugita,
\& Okamura (1993). This parameter closely tracks bulge-to-disk ratios
and has been shown to provide a quantitative substitute for more
orthodox visual classifications of position along the Hubble Sequence
(Abraham et al.\ 1996).  The concentration parameter is simply the
flux ratio between an inner and a outer ellipse. The outer ellipse is
selected using second order moments obtained by a 2 $\sigma$ cut above
the sky noise, $\sigma$.  The inner ellipse is calculated by a similar
moment analysis, but at a radius of only 30\% of the outer ellipse.  

Assuming that the galaxy is intrinsically axisymmetric at large radii,
the axis ratio of the outer ellipse can be used to define the galaxy
inclination, which is also important for determining the bar shape
($y$) axis. Following Abraham \& Merrifield (2000) we define the bar
strength as;
\begin{equation}
f_{bar} = \frac{2}{\pi}\left[
               \arctan\left (\frac{b}{a}\right)_{bar}^{-\frac{1}{2}} 
             - \arctan\left (\frac{b}{a}\right)_{bar}^{+\frac{1}{2}}
                       \right],
\end{equation}
where $(\frac{b}{a})_{bar}$ is the intrinsic axis ratio of the 
putative bar,
calculated from its apparent axis ratio and the galaxy's inclination
using equation~(2) of Abraham et al.\ (1999).  To make sure that we do
not miss any small high surface brightness or large low surface
brightness bars, the whole range of radii is searched starting with
the outer isophote (defined as for C) and working inwards. The inner
ellipse is then defined as the isophote with the minimum 
$(\frac{b}{a})_{bar}$ (i.e.\ the maximum value of $f_{\rm bar}$).  
Figure~\ref{ellipse} shows several examples of the ellipse fits.

In order to discriminate between barred and unbarred galaxies, Abraham et al. (1999) proposed that systems with $(\frac{b}{a})_{bar}^{2} < 0.5$ could be classified as barred.  This corresponds to $f_{bar} < 0.11$, which we will use to determine the fraction of barred galaxies in the OSUBSGS. It should be stressed that this cutoff does not represent any underlying physical process within spiral galaxies, but merely provides a useful visual criterion for defining a galaxy as barred.

\section{Results}\label{sec:results}

\begin{figure}
\begin{minipage}[b]{\linewidth}
\psfig{file=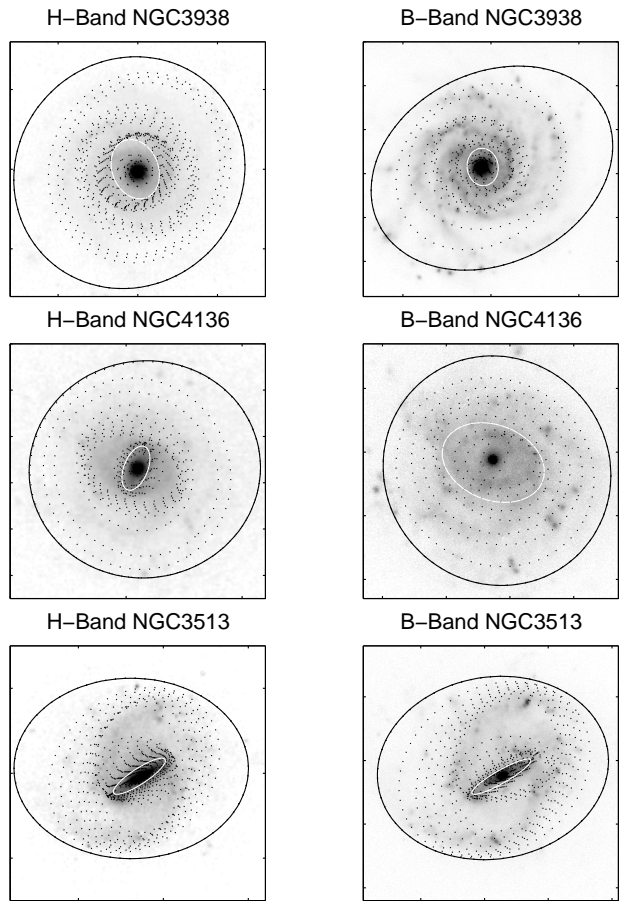,angle=0,width=\linewidth}
\end{minipage}
\caption{Ellipse fits for H and B band images. The solid black line is
the outer ellipse used to determine the galaxy inclination, and for
calculations of C and $f_{bar}$. The white ellipse is the inner
ellipse used for measuring $f_{bar}$. The dashed lines are isophotes
fitted to different intensity slices separated by $1\sigma$ in the sky
noise.}
\label{ellipse}
\end{figure}

\begin{figure*}
\begin{minipage}[b]{.5\linewidth}
\psfig{file=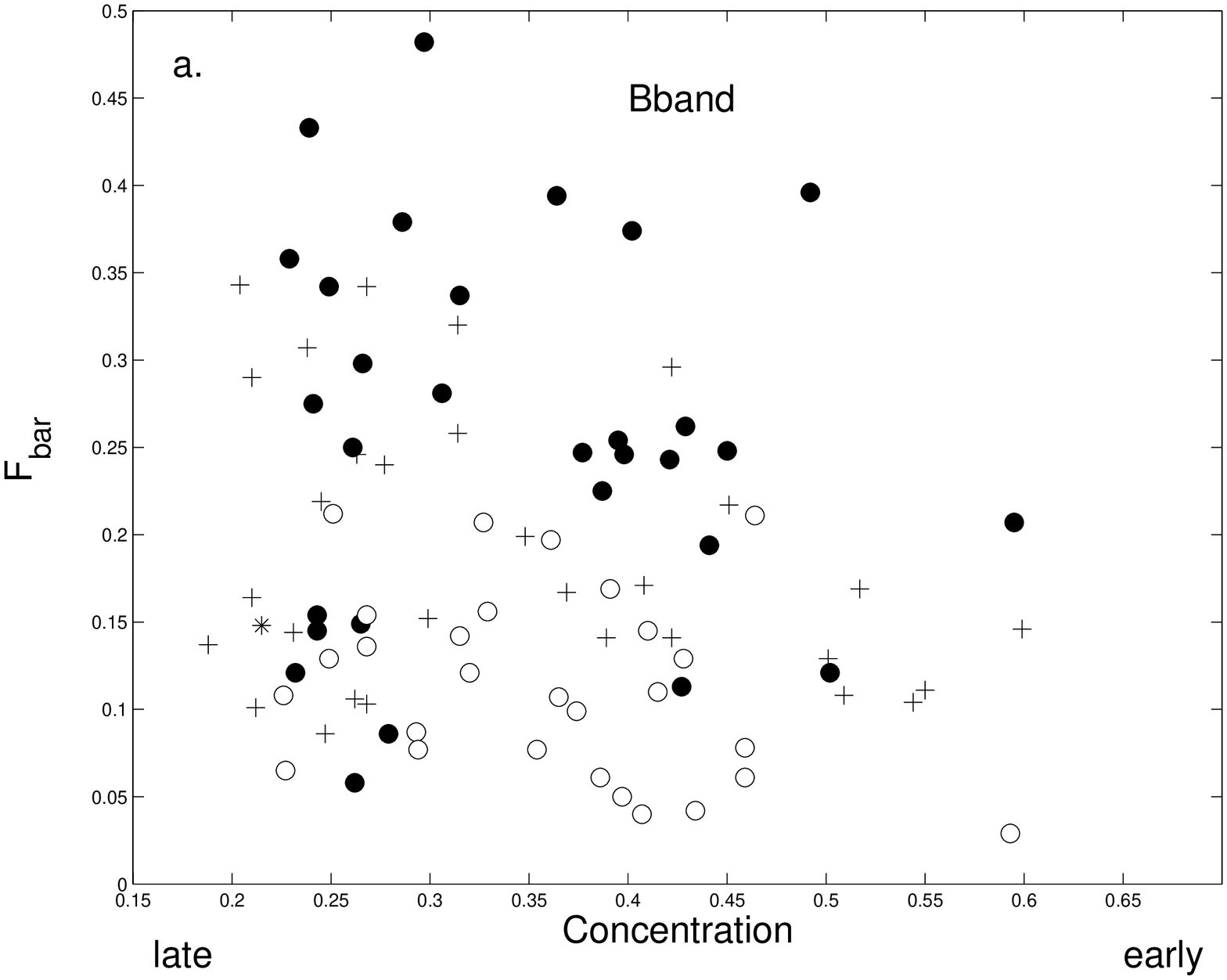,angle=270,width=\linewidth}
\end{minipage}\hfill
\begin{minipage}[b]{.5\linewidth}
\psfig{file=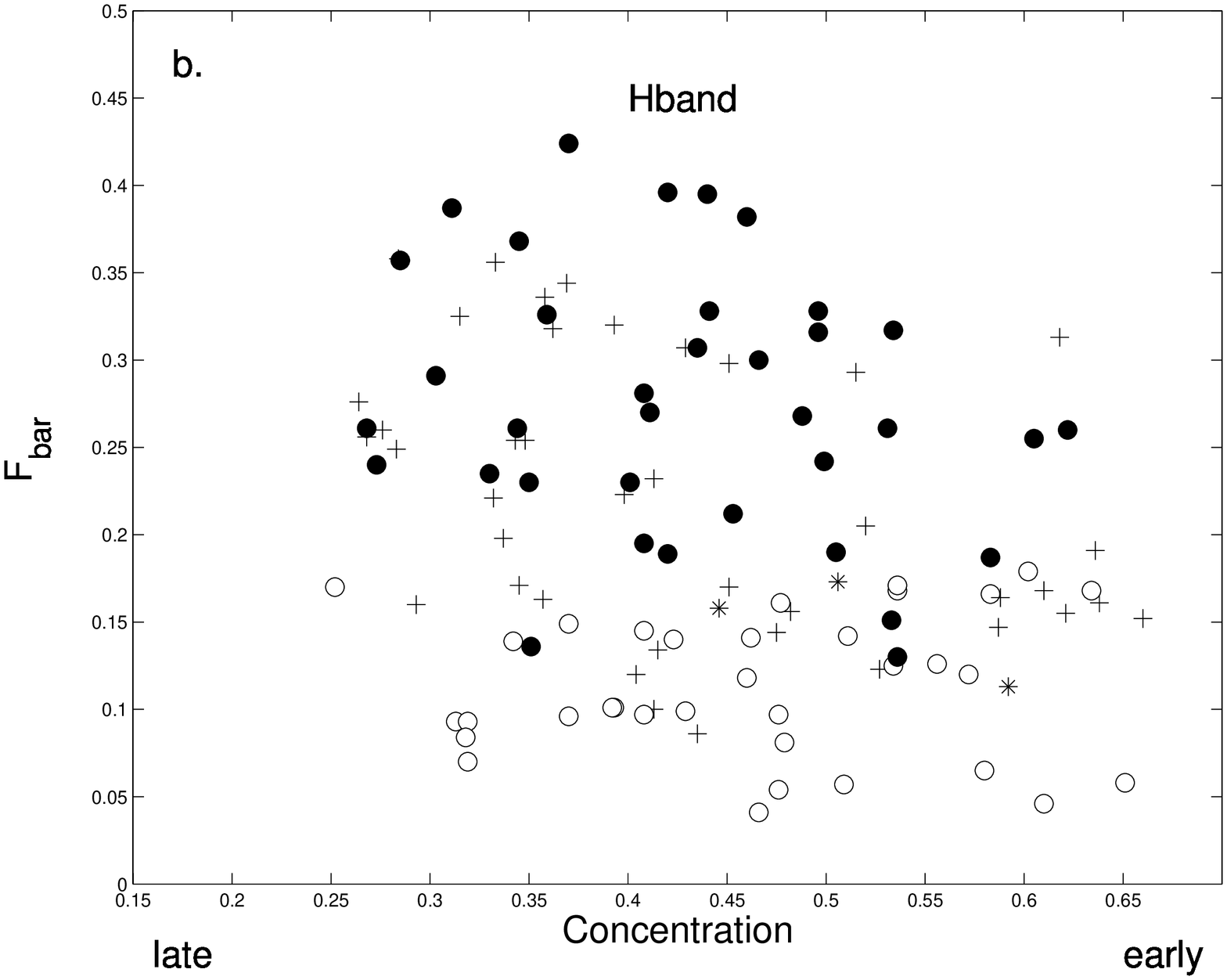,angle=270,width=\linewidth}
\end{minipage}
\caption{Distribution of optical and near infrared samples
in ``Hubble Space''. SA galaxies are represented by circles, SAB
galaxies by crosses, SB galaxies by filled circles, and unclassified
galaxies by asterisks.}
\label{galdis}
\end{figure*}

Figure~\ref{galdis} displays the distribution of the optical and near infrared samples in the two dimensional Hubble Space.  A slight
anticorrelation is observed between $C$ and $f_{\rm bar}$ in both
infrared and optical bands.  As discussed in Abraham \& Merrifield
(2000), such a trend occurs naturally as the presence of a strong
bulge (and hence large value of $C$) will tend to wash out any bar
signature, whereas in galaxies with weak bulges even quite modest bars
will still be strongly detected.

\begin{figure}
\begin{minipage}[b]{.95\linewidth}
\psfig{file=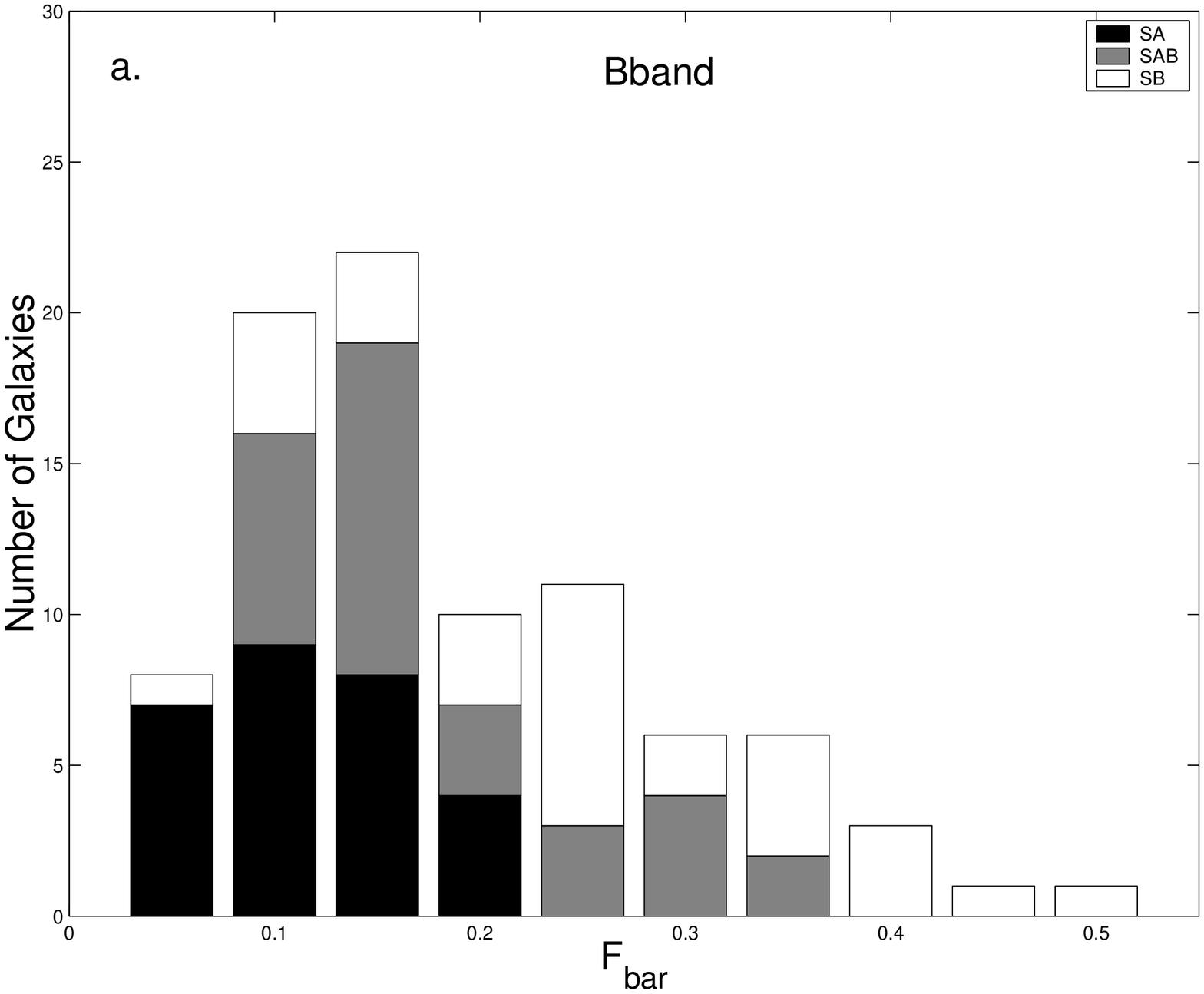,angle=270,width=\linewidth}
\psfig{file=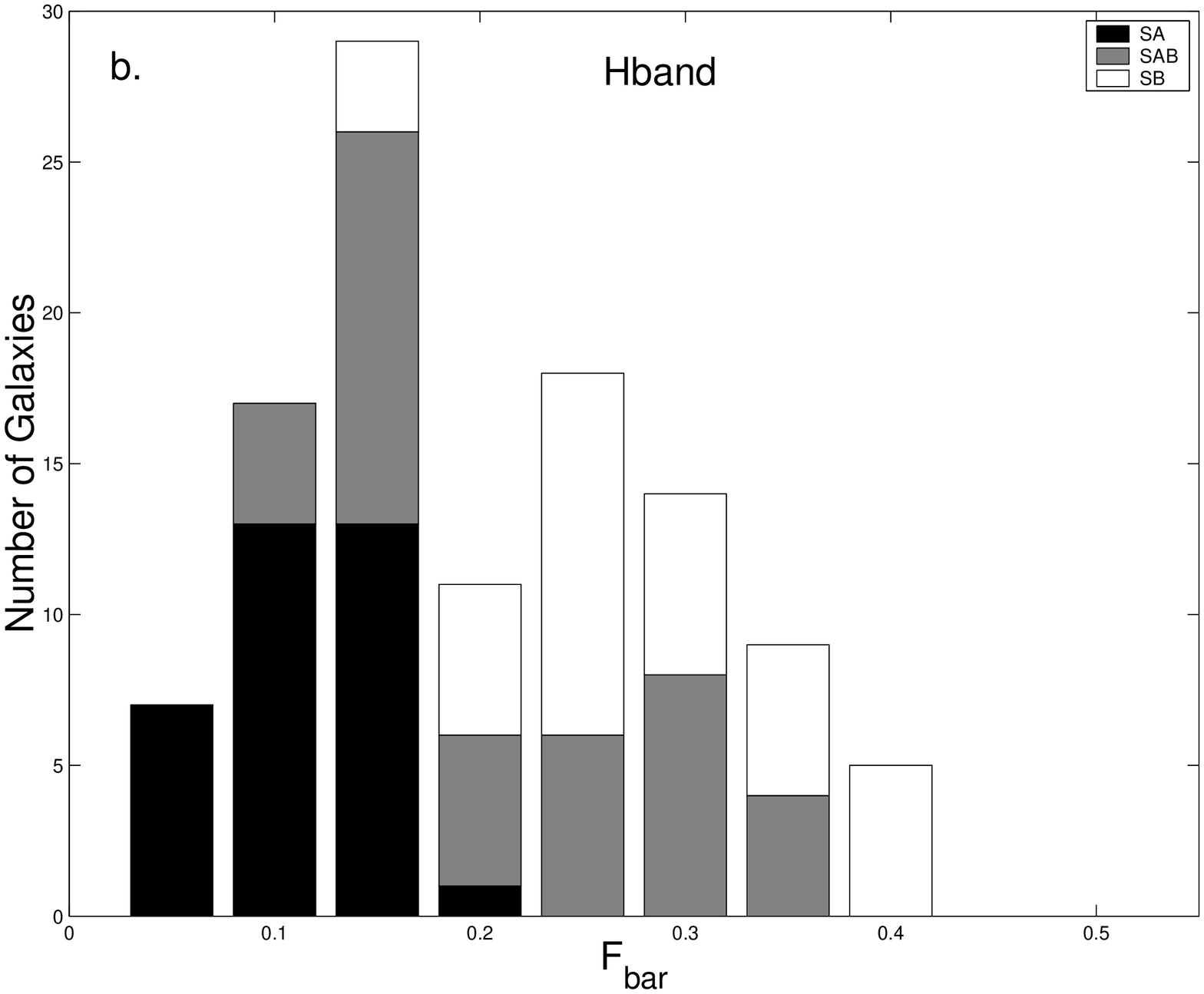,angle=270,width=\linewidth}
\end{minipage}
\caption{Histogram of bar shape distribution in (a) Optical and (b.) Near Infrared.}
\label{histfbar}
\end{figure}

In terms of their qualitative bar classifications, the galaxies are
reasonably well divided in both optical and near infrared Hubble Space, with SA
galaxies distributed across the bottom, SB galaxies across the top,
and SAB galaxies bridging the gap between the two.  The bimodal split
between barred and unbarred galaxies is far less apparent in these
figures than was found by Abraham \& Merrifield (2000).  However,
there is still evidence for such a split even in this more objectively
defined sample.  The histograms of $f_{\rm bar}$ in
Fig.~\ref{histfbar} show a dip in the number of galaxies at $f_{\rm
bar} \sim 0.2$, the same value as the gap found in Abraham \&
Merrifield's (2000) analysis of the Frei sample.  The dip is at a
greater significance level ($\sim 95\%$) in the infrared data than in
the optical, which suggests that the optical bar shapes are more
compromised by the effects of star formation, dust obscuration, etc.

This impression is borne out by the fact that there is a tighter
grouping of the visual classifications, particularly the SB galaxies,
in the infrared data than in the optical (despite the fact that the
visual classifications were based on optical images).  The $5\%$ of cases where the visual and automated classifications disagree strongly were inspected individually.  In almost all 
cases the discrepancy arises because the automated fit is being compromised by bright star-forming features.  In some cases, these features are localised
regions, but in others (such as NGC4136, see Fig.~\ref{ellipse}) the fit is
drawn to tightly wrapped spiral arms, which are identified as the
strongest non-circular features, rather than the bar that we seek.  It should be noted that for galaxies with two or more bars, the automated process would pick out the most elliptical bar, not necessarily the primary outer bar (although no such cases were found in this sample).  We could have intervened in the bar analysis to prevent such misfits, but such subjective judgements are contrary to the ethos of this approach, as they could easily introduce subjective biases in the analysis.

There are thus at least two non-physical sources of scatter in the
distribution of $f_{\rm bar}$.  First, the intrinsic limitations of
the automated fitting process (particularly at optical wavelengths)
mean that there is an uncertainty introduced in the derived value of
bar shape.  Second, the correlation between $C$ and $f_{\rm bar}$
means that a projection of this two-dimensional space onto the 
$f_{\rm bar}$ axis will tend to smear out intrinsically sharp features in
Hubble space when viewed in one dimension.  Hence, the true divide
between barred and unbarred galaxies is likely to be rather more
dramatic than we see in Fig.~\ref{histfbar}.

In order to compare the fraction of barred galaxies in the optical and infrared, only the 72 galaxies for which images in both bands were available were used.  We find that $74\%$ of galaxies are barred in the optical, with this fraction increasing to $79\%$ in the near infrared. Visual classification of the entire OSUBSGS sample, preformed by Eskridge et al.\ (2000) found only $72\%$ of galaxies to be either strongly or weakly barred in the \textit{H}-band, while only 64\% were classified as barred in the RC3 (\textit{B}-band).   Overall there is a good correlation between optical and NIR bar shape.  Obscuration from dust and star-formation did cause some late-type galaxies to appear more barred in the NIR than the optical.  A few galaxies did appear to be significatly more barred in the optical than NIR, as isophotes fitted around HII regions can occasionally give the appearance of a bar.

Automated classification finds a slightly higher fraction of barred galaxies in both bands than visual classification.  This could mean that smaller bars are occasionally overlooked when visual classifications are made. However the difference in fractions in this case is probably due to the value of $f_{bar}$ selected to separate barred from unbarred galaxies.  Indeed, if we increase the proposed value of $f_{bar} < 0.11$ to $f_{bar} < 0.14$, we find that 61\% of galaxies are barred in the \textit{B}-band and 71\% in the \textit{H}-band.  These fractions are in good agreement with those found by Eskridge et al.\ (2000).

\begin{figure}
\begin{minipage}[b]{\linewidth}
\psfig{file=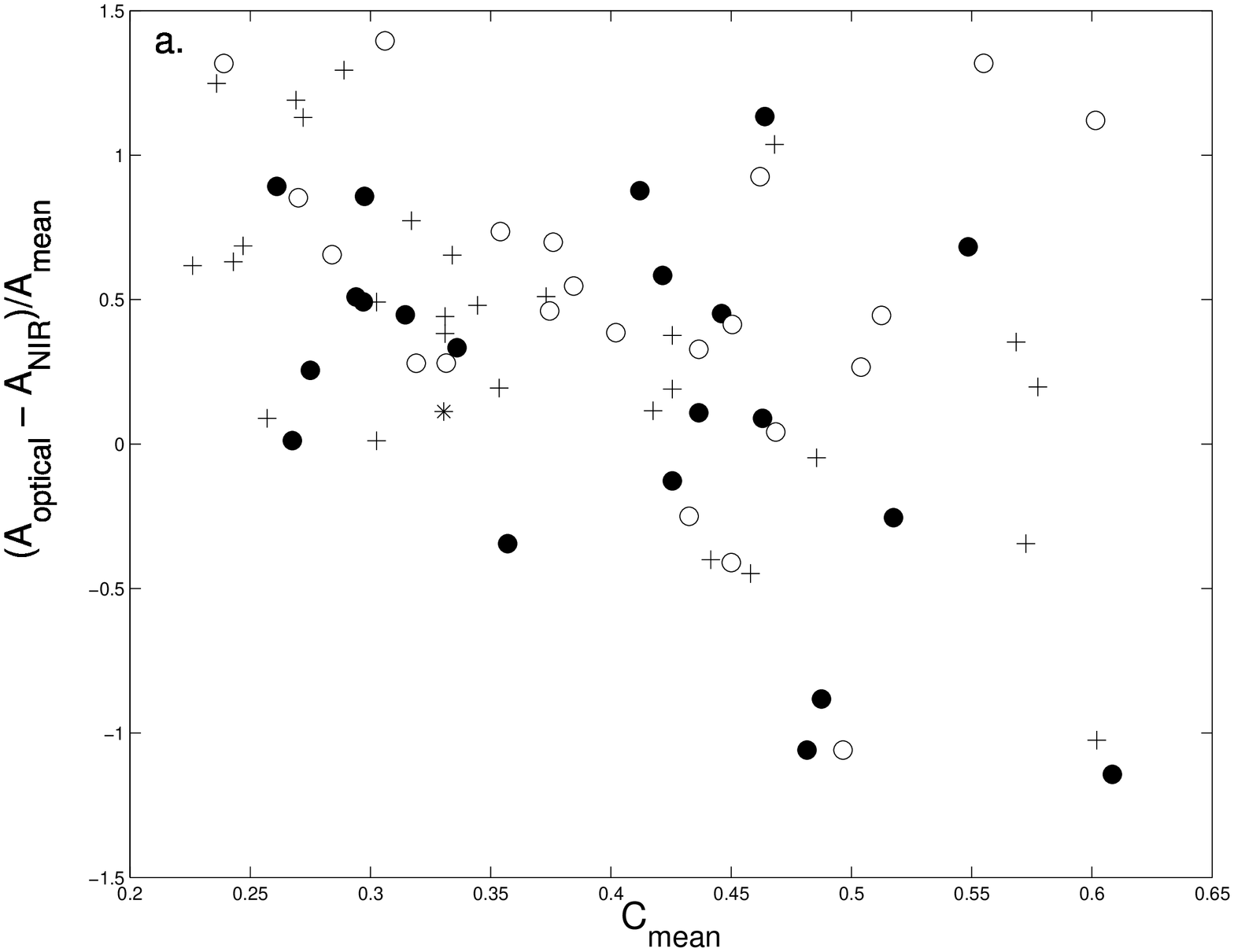,angle=270,width=\linewidth}
\psfig{file=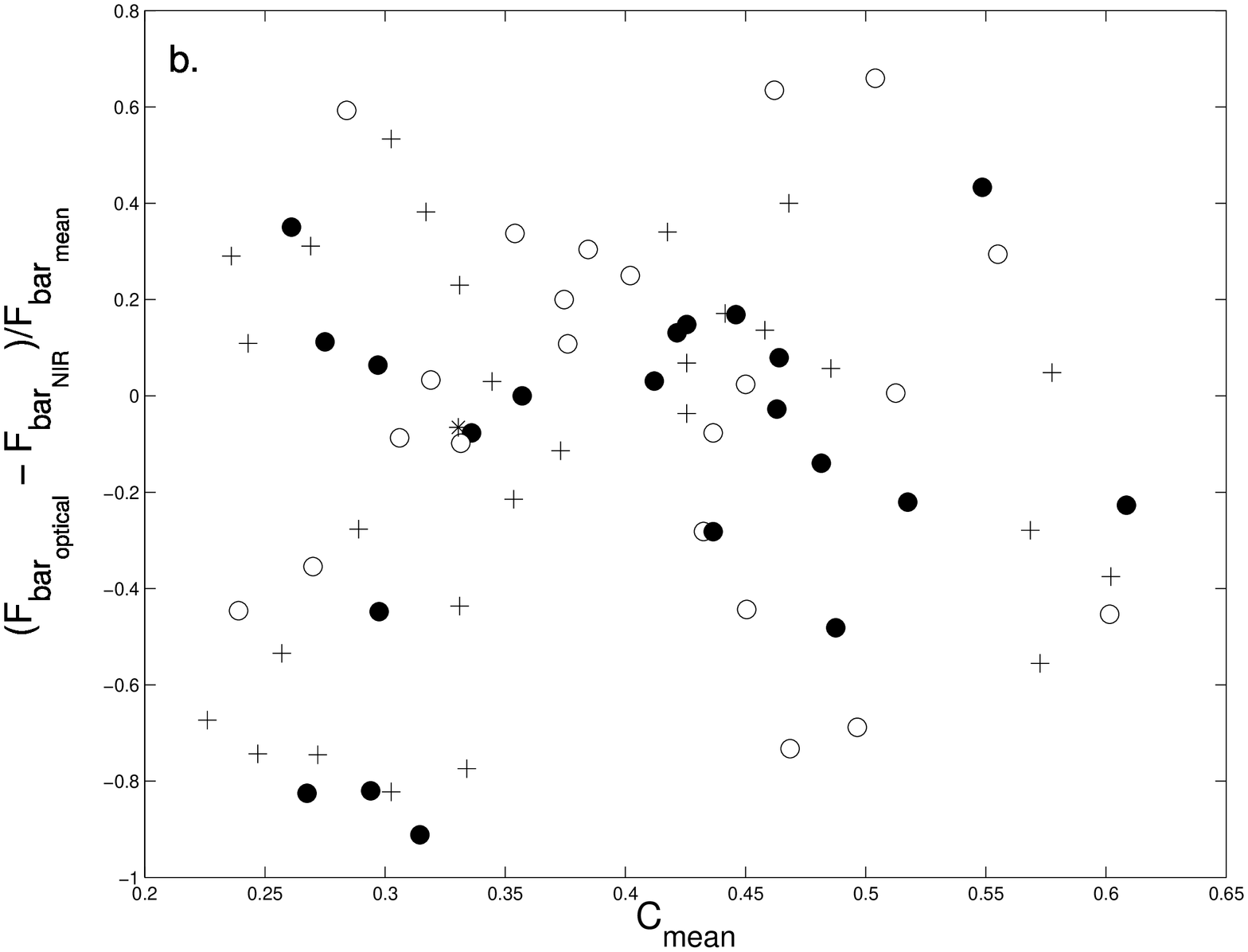,angle=270,width=\linewidth}
\psfig{file=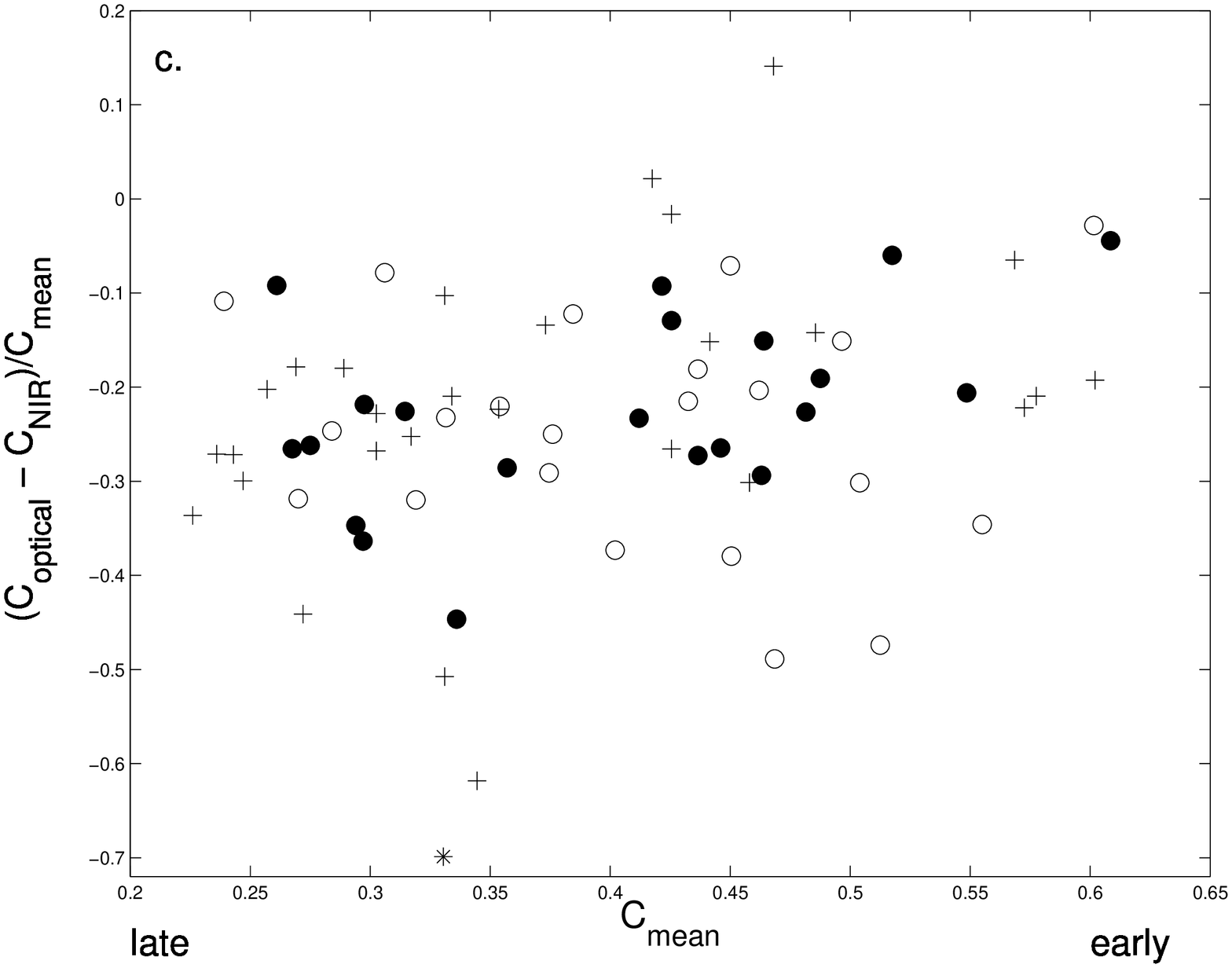,angle=270,width=\linewidth}
\end{minipage}
\caption{Variation between the optical and infrared data of (a.) the
asymmetry index, (b.) the bar shape parameter, and (c.) the central
concentration parameter, plotted as a function of the galaxy's
concentration.  SA galaxies are represented by circles, SAB galaxies
by crosses, SB galaxies by filled circles, and unclassified galaxies by
asterisks.}
\label{diff}
\end{figure}

Having established that the optical properties of galaxies introduce
more dispersion in their objective classifications than occurs in the
infrared, we now look briefly at possible systematic differences
between the two data sets.  Figure~\ref{diff} shows the fractional
variation in asymmetry, bar shape and central concentration between
the optical and infrared data as a function of mean concentration (the
average of the concentrations in the two bands).  There is a clear
trend in asymmetry in the sense that although high concentration
(early type) galaxies have similar asymmetries, the low concentration
(late type) galaxies are systematically more asymmetric in the optical
than the infrared.  Patchy, asymmetric star formation occurs more
commonly in later-type galaxies, so this result again reflects the
fact that optical data are more affected by star formation than
infrared observations.  Bar shape does not seem to be
systematically affected in the same way.  Although patchy star
formation increases the scatter in bar shape in low concentration
galaxies, it only marginally lowers the bar shapes in the optical
compared to their infrared values.  Concentration, on the other hand,
does change systematically between the optical and infrared in the
sense that galaxies are on average more centrally concentrated in the
infrared.  This variation could be an indicator of opacity in the
central regions of these galaxies, which suppresses the optical
emission -- as recently discussed in White, Keel \& Conselice (2000),
this remains a controversial issue.  However, it could also simply
reflect the star formation in the disks of these galaxies, which will
augment the optical emission at large radii, lowering the
concentration observed in this band.

\section{Discussion}\label{sec:disc}

This study has placed a well-defined sample of spiral galaxies in the
Hubble Space of early-to-late type and bar shape.  The main
conclusion is that there are significant signs of bimodality in the
bar shape parameter, similar to those found by Abraham \&
Merrifield's (2000).  Since the current analysis is based on a sample
with more objective selection criteria, this bimodality is unlikely to
be attributable to selection effects.  This therefore suggests that
barred and unbarred galaxies exist as distinct populations, rather
than as the extremes of a continuum.  

Barred and unbarred galaxies are remarkably similar in 
all their other properties such as size, luminosity, Tully-Fisher 
relation, etc, (e.g.\ Debattista \& Sellwood 2000).  Therefore it seems unlikely that barred and unbarred galaxies are two separate types of system with no evolution between the two.  Further, a variety of mechanisms have shown up
in numerical simulations that may cause bars to form and dissolve in
real disk galaxies (e.g.\ Athanassoula 2002).  Thus, there are good 
reasons to believe that evolution should occur between these two
populations.  The existence of a gap between them then implies that
such evolution must occur on a timescale that is short compared to the
cosmological lifetimes of the galaxies (just as the Hertzsprung gap in
the stellar colour-magnitude diagram indicates the haste with which
stars evolve through this region of parameter space).  

The comparison between optical and near-infrared morphologies is also
instructive.  Some measures of galaxy morphology, such as central concentration and asymmetry, vary systematically between the optical and near infrared.  However the bar shape parameter remains approximately the same, albeit with greater scatter in the optical due to localised star formation, etc. This discovery has an important implication for studies of bar shape as
a function of redshift, which indicate that barred galaxies are
extremely rare for $z \geq 0.5$ (Abraham et al. 1999; Van den Bergh et
al. 2000).  One possible explanation for this rarity is that it might
be attributed to bandshifting effect, since the \textit{I}-band images used in
these studies correspond to \textit{B}-band rest-frame emission for $z \sim
0.8$.  Thus, if bars were systematically weaker in the B band than at
red wavelengths, then one would expect to see such a decrease in bar
fraction with redshift. This matter is discussed in greater detail 
by Van den Bergh et al. (2002).  Since there is no such systematic variation
in bar shape with waveband, this explanation now seems unlikely; it
would appear that bars are intrinsically rarer at higher redshifts.

These conclusions have been drawn from a relatively small sample of
galaxies, necessarily limiting their confidence.  However, with the
advent of large surveys like the Sloan Digital Sky Survey (York et
al.\ 2000), much larger samples of galaxies are becoming available.
Since the morphological parameters used in this analysis are
reasonably robust, they can be applied to rather more distant galaxies
than traditional classifications, opening up a larger region of space
to such studies, and hence allowing much larger samples to be
gathered.  With these larger samples, objective morphological analysis
should provide definitive answers to many of the questions where this
study has only been able to scratch the surface.

\section{acknowledgements}
This work made use of data from the Ohio State University Bright
Spiral Galaxy Survey, which was funded by grants AST-9217716 and
AST-9617006 from the United States National Science Foundation,
with additional support from the Ohio State University.

\bsp

\label{lastpage}

\end{document}